\documentclass[aoas,preprint]{imsart}
\RequirePackage[OT1]{fontenc}
\usepackage{amsthm,amsmath,natbib}
\usepackage{float}

\usepackage[pdftex]{graphics,color}
\usepackage{amsmath}
\usepackage{amssymb}
\usepackage{amsmath}
\usepackage{bm}
\usepackage{siunitx}
\usepackage{amssymb}
\usepackage{pxfonts}
\usepackage{graphicx}
\usepackage{dcolumn} 
\usepackage{array}
\usepackage{url}
\usepackage{rotating} 
\newcommand{\betadist}{{\mbox{beta}}}

\newcommand{\Bern}{{\mbox{Bern}}}
\newcommand{\logit}{{\mbox{logit}}}

\newcommand{\Var}{{\mbox{Var}}}

\newcommand{\EEE}{{\mbox{E}}}

\newcommand{\NNN}{{\mbox{N}}}

\newcommand{\bfPsi}{{\mathbf{\Psi}}}
\newcommand{\bfzero}{{\mathbf{0}}}
\newcommand{\bfSigma}{{\mathbf{\Sigma}}}

\startlocaldefs
\numberwithin{equation}{section}
\theoremstyle{plain}

\endlocaldefs

\begin{document}



\begin{frontmatter}
	\title{Measuring Within and Between Group Inequality in Infant Mortality Over Time: A Bayesian Approach with Application to India\thanksref{T1}}
	\thankstext{T1}{Antonio P. Ramos is Postdoctoral Fellow, Department of Biostatistics, Fielding School of Public Health, and the California Center for Population Research, University of California, Los Angeles, Los Angeles, CA, 90095-1772 (E-mail: tomramos@ucla.edu). Robert E. Weiss is Professor, Department of Biostatistics, Fielding School of Public Health, University of California, Los Angeles, Los Angeles, CA, 90095-1772 (E-mail: robweiss@ucla.edu). Ramos was supported by the Eunice Kennedy Shriver National Institute of Child Health and Human Development, research grant K99HD088727.}
	
	\begin{aug}
		\author{\fnms{Antonio P.} \snm{Ramos}}
		\and
		\author{\fnms{Robert E.} \snm{Weiss}}
	\end{aug}
	
	\begin{abstract}
	Most studies on inequality in infant and child mortality compare average mortality rates between large groups of births, for example, comparing births from different countries, income groups, ethnicities, or different times. These studies do not measure within-group disparities. The few studies that have measured within-group variability in infant and child mortality have used tools from the income inequality literature, such as Gini indices. We show that the latter are inappropriate for infant and child mortality. We develop novel tools that are appropriate for analyzing infant and child mortality inequality, including inequality measures, covariate adjustments, and ANOVA methods. We illustrate how to handle uncertainty about complex inference targets, including ensembles of probabilities and kernel density estimates. We illustrate our methodology using a large data set from India, where we estimate infant and child mortality risk for over 400,000 births using a Bayesian hierarchical model. We show that most of the variance in mortality risk exists within groups of births, not between them, and thus that within-group mortality needs to be taken into account when assessing inequality in infant and child mortality. Our approach has broad applicability to many health indicators.
	\end{abstract}
	
	\begin{keyword}
		\kwd{Complex Inference Targets}
		\kwd{Compositional adjustments}
		\kwd{Demographic methods}
		\kwd{Health inequality}
		\kwd{Income inequality}
	\end{keyword}
	
\end{frontmatter}


\section{Introduction}

\noindent Inequality in neonatal, infant, and child mortality are fundamental dimensions of social inequality \citep{Moser:2005,Stuckler:2010}. They are usually defined by differences in average mortality rates between groups of births characterized by a single demographic, such as race, gender, income or country. In high income countries, national averages of child mortality are less than 10 deaths per thousand births, while these rates can be over 200 deaths per thousand births in low income countries. Within-country disparities in mortality can be just as large, particularly across groups defined by wealth and income \citep{Victora:2003,Sastry:2004,Wagstaff:2000} and race and ethnicity \citep{Brockerhoff:2000,Antai:2011,Jankowska:2013}.

If births within a group have similar mortality risk, quantification of between-group disparities based on group averages is sufficient to characterize disparities in infant and child mortality in human populations. If births within a single group have widely varying underlying mortality risk, then within-group variation in mortality risk can be higher than between-group variation. If that is the case, then comparing groups based on their averages only provide us with a partial picture of the total inequality in mortality within societies. Similarly, policy interventions that aim at reducing disparities in infant and child mortality usually divide society into groups and target the groups with the highest average mortality risk. However, large variability in mortality risk  within a group can lead to inefficient policies, as high risk births may exist in large numbers in many if not all groups. For example, \cite{Ramos:2018} shows that targeting on poverty alone leads to program inefficiency because \emph{most} high risk births are not found on the poorest group. 

Previous literature on life span variability suggests that within-group variability is larger than between-group variability. \cite{Edwards:2005} find  heterogeneous trends over time in adult life span variability across industrialized nations, despite convergency in average levels. This finding was further confirmed in 180 nations \citep{Edwards:2011}. Using a new decomposition method, \cite{Nau:2012} found life expectancy is more variable in United States than in Sweden. Similarly, Hispanics have less variability in life expectancy than Whites \citep{Lariscy:2016}, who have less variability than Blacks \citep{Glenn:2014}. We study infant mortality in this paper which are distinct from life span and thus requires separate analysis and understanding.

Previous work on the measurement of inequality in infant and child mortality has borrowed inequality measures from the income inequality literature \citep{Gakidou:2002, Gakidou:2003, Murray:2001} and these measures have been implemented by international organizations \citep{WHO:2000,India:2012}. However, it is not clear that income inequality measures are appropriate for studying mortality risks. Mortality is measured on the probability scale and thus is bounded inside the $[0,1]$ interval. Income, however, is usually defined on the positive real line or, sometimes, on the whole real line. Life expectancy is also similar to income because it has support on $[0,\infty)$. Some work on variability in life expectancy has also used measures from the income inequality literature \citep{Peltzman:2009}. However, properties that are appropriate for measures defined on the real line, such as scale invariance, are not appropriate for probabilities. Thus while it is clear that within-group variability in infant and child mortality should be quantified, it is unclear how to do so.

Demographic decomposition techniques, such as \cite{Kitagawa:1955} or \cite{Oxaaca:1973}, have been applied to explain \emph{average} differences in infant and child mortality and other health indicators between populations. However, two populations with the same average rates of infant and child mortality can have very different distributions of mortality risk \citep{Gakidou:2002}. Sophisticated techniques have also been developed for variance decomposition in inequality in life expectancy \citep{Edwards:2005,Edwards:2011,Nau:2012,Glenn:2014, Lariscy:2016}. However, these methods only decompose variance between two populations or time points and are not designed to be applied to other aspects of distributional changes. Thus, it is important to develop a framework that quantifies and explains differences in distributions of mortality risk beyond mean and variance changes and that are appropriate to mortality data. Currently no methodology measures and explains differences in the distribution of mortality risk across populations in  a manner similar to decompositional methods in labor economics \citep{Fortin:2011} or relative distribution methods \citep{Handcock:1999}.

Mortality risk is not an observable quantity, however. We only observe that births are still alive or not at a certain age. To measure and explain inequalities in infant and child mortality across births we propose to estimate birth risk using a statistical model. Only then can we proceed to the inequality analysis. To make proper inferences we will need to propagate uncertainty from the estimation stage to the analysis of inequality stage.

In this paper we develop new methods to analyze the distribution of mortality risk with the objective of measuring and explaining health inequality within and across populations. We demonstrate our methodology using two waves of the Demographic and Health Surveys (DHS) with information on more than 400,000 births from India. We estimate the underlying infant mortality risk with associated uncertainty and complex posterior covariance structure using Bayesian hierarchical logistic regression fit via Markov chain Monte Carlo (MCMC). Bayesian estimation and MCMC methods can handle complex models and facilitate the propagation of model and estimation uncertainty into subsequent analysis of inequality. Hierarchical models are a natural choice in our context because cases in our data set are nested in larger sampling units: births are nested in mothers, mothers are nested within sampling clusters, which are nested in districts and districts in states. Our inference target is the empirical distribution of mortality risk both marginally and as a function of covariates.

We make a number of contributions with broader applications to the existing literature on inequality in infant and child mortality. First, we show that the usual practice of measuring inequality in infant and child mortality using measures from the income inequality is not appropriate. Thus we suggest that to investigate infant and child mortality one needs measures that are specifically developed for mortality. 

Second, we show that proper assessment of inequality requires consideration of within-group variability. We introduce methods that are appropriate for quantifying inequality in infant and child mortality within and between groups of births. We introduce summary measures that quantify the overall difference between two distributions. We develop adjustments that parsimoniously summarize differences between distributions of mortality risk. We extend covariate adjustment methods from the demographic literature by expanding the types of comparisons. We propose comparisons between distributions, while most research in demography and public health only compares means or other summary measures. Our methods allow us to separate the impacts of changes in population composition from changes in the covariate-outcome relationship. This makes it possible to answer questions such as how the distribution of mortality risk in 1995 would have looked if there had been no changes in the distribution of maternal age since 1975. We also develop ANOVA methods applied after Bayesian model fitting to quantify within-and-between group variance in mortality risk. \cite{Gakidou:2002} have suggested that most of the variance in mortality risk occurs within groups of births, not between them. We develop methodology to formally demonstrate the truth of this hypothesis. Our methodology directly allows for complex inference targets while allowing for measurement of uncertainty. Inequality measures and ANOVA are examples of new complex inferential targets that benefit from Bayesian inference.

Third, our analysis uncovers several new patterns of inequality in infant and child mortality in India. Inequality in infant and child mortality in India has been widely studied \citep{Bhattacharya:2012,Singh:2011,De:2013,Jain:2013,Kumar:2014,India:2012}. The level of aggregation varies in these studies (states, districts, etc.), but great inequality has been documented. However, inequality in infant and child mortality  at the individual birth level has not been investigated previously. We show that looking at average mortality rates among socioeconomic groups provides an incomplete picture of infant and child mortality inequality in India. For example, demographic groups defined by caste, maternal age at the birth of the child and religion explain less than 4\% of all variance in infant and child mortality risk. Groups defined by quintiles of wealth explain between 7\% and 10\% of all variance. District has the highest explanatory power but still explains less than 20\% of all variance in infant and child mortality risk. These patterns are consistent over time and suggest that previous analyses have largely overlooked variability in infant and child mortality risk within groups of births. We show that differences over time in the distribution of infant and child mortality risk can be summarized by a simple multiplicative shift. And, finally, we show that changes over time in covariate distributions only account for a small fraction of the changes over time in infant and child mortality risk.

The paper is organized as follows. In section \ref{problems} we show that income inequality measures are not appropriate for studying infant and child Mortality. Section \ref{analyzing} suggests summary measures and adjustments that are appropriate for assessing inequalities in infant and child Mortality. Section \ref{covariates} introduces covariate adjustment methods. Second \ref{accounting} discuss the uncertainty quantification while evaluating inequality. Section \ref{modeling} illustrate our methodology using data from India. Section \ref{discussion} discusses some of the methodological and policy implications of our findings. Details of the data and the complete model we use for estimating infant mortality risk are presented in the appendix.

\section{Problems with applying measures from the income inequality literature to mortality data}

\label{problems}

Most measures of income inequality have a common form \citep{Firebaugh:2002}. For a population of $i = 1, \ldots, n$ births with mortality risk $\pi_{i}$, income inequality measures take the form
\begin{align}
\text{ineq} & = \frac{1}{n}\sum_{i=1}^{n}f\left(\frac{\pi_{i}}{\mu}\right),
\label{inequality}
\end{align}
where $\mu = n^{-1}\sum_{i}^{n}\pi_{i}$ is the average risk $\mu$ in the population and $f(\cdot)$ is a function to be specified. Define $r_{i} = \pi_{i}/\mu$ which is the ratio of the $i$ birth mortality risk to the population average risk; inequality measures \eqref{inequality} are functions of $r_i$. Three popular measures of inequality are the squared coefficient of variance, Theil indexes, and variance of the logs
\begin{align}
f_{1}(r_{i}) & = (r_{i}-1)^{2} \label{cov} \\
f_{2}(r_{i}) & = r_{i} \log(r_{i}) \label{varlog} \\
f_{3}(r_{i}) & = \left\{\log(r_{i})-\frac{1}{n}\sum_{i=1}^{n}\log(r_{i})\right\}^{2} \label{theil}
\end{align}
and the closely related Gini index
\begin{align}
\text{Gini} & =  \frac{1}{2n^2}\sum_{i=1}^{n}\sum_{j=1}^{n}\left| r_{i} - r_j \right|. \label{gini}
\end{align}

There are at least three problems in applying \eqref{inequality} or \eqref{gini} to mortality data. First, in \eqref{inequality} or \eqref{gini} the average level of income $\mu$ of the population under study is in the denominator. This is not a problem for income, as average income is usually far from zero and income inequality measures are designed to be the same whatever currency or units income is measured in or if income is doubled or halved across the board. Applied to mortality data, these indices divide the expected mortality risk for each birth, $\pi_{i}$, by the average mortality risk, $\mu$. However, as $\mu$ approaches zero, the ratio $r_{i}$ becomes large for fixed $\pi_{i}$, which increases the value of the inequality index. Thus a numerical problem may give a false impression that Infant and child mortality inequality is increasing even as infant and child Mortality rates tend towards zero.

The second problem is that probabilities are bounded by the $(0,1)$ interval, unlike income which is defined on the real line. For income inequality, for a set of incomes $s_i$, $i = 1, \ldots, n$ and constant $c$, for all measures \eqref{inequality} -- \eqref{gini} $\mbox{ineq}(s_1, \ldots, s_n) = \mbox{ineq}(cs_1, \ldots, cs_n)$, income inequality measures remain unchanged if we multiply every income by $c$. Scale-invariance, which is sensible for income data does not readily translate to mortality risk. The probability scale imposes serious constraints on $c$ as the distribution of $c\pi_i$ also needs to be in the $(0,1)$ interval; at best $c$ must be less than 1 over the maximum of the $\pi_i$s; for probabilities that cover the full range $(0,1)$, only choices of $0<c<1$ are acceptable.

The third problem with inequality measures from the income inequality literature is that the inequality indices are not consistent with basic intuition about justice and health. Inequality measures on probability distributions should lead to the same conclusion whether we are measuring probability of mortality, $\pi_{i}$, or probability of survival, $1-\pi_{i}$. A country with high inequality in infant mortality has equally high inequality in infant survival. Thus we assert a symmetry property in the evaluation of inequality on probability distributions

\begin{itemize}
\item \textbf{Symmetry:} Suppose $Z_1$ and $Z_2$ are two probability distributions on $[0,1]$. Then comparing inequality between $Z_1$ and $Z_2$ and comparing inequality between $1-Z_1$ and $1-Z_2$ should produce the same conclusion.
\end{itemize}

We use simulated data from a series of beta distributions with decreasing means that resemble mortality risk distributions to illustrate how inequality measures behave. Table 1 presents results for four $\betadist(\alpha,\beta)$ distributions, where $\beta$ is fixed at $10$ for all four distributions and $\alpha$ decreases from $1$ to $.5$, $.3$ and $.01$. As $\alpha$ decreases, the mean $\alpha / (\alpha + \beta)$ and standard deviation $\alpha\beta/[(\alpha + \beta)^2 (1 + \alpha + \beta)]$ of mortality risk decrease. The mean gives the same interpretation whether we look at survival or mortality: survival is increasing and mortality is decreasing top to bottom; we interpret this as inequality is also decreasing from top to bottom. The standard deviation is the same whether we look at mortality or survival and suggests that inequality is decreasing from top to bottom. The coefficient of variation gives opposite stories about inequality for mortality and survival as do the Gini and Theil inequality indices: increasing from top to bottom for mortality, but decreasing for survival.

Income inequality measures such as CV, Gini or Theil do not satisfy the symmetry property and therefore are not suited to assess inequality in probability distributions. Income distributions are fundamentally different from probability distributions. The idea of scale invariance is neither necessary nor appropriate for measuring inequality in mortality because probabilities do not scale. In contrast, the mean or variance appear to be possible measures of inequality for distributions of infant and child mortality risk.

We now look at other methodologies to measure and describe inequality in infant and child mortality risk.

\section{Analyzing Inequality in Early Life Mortality}

\label{analyzing}

In this section we develop appropriate tools to study inequality in mortality data. We begin with definitions and notation. Consider two populations we wish to compare, a reference population $j=0$, and a comparison population $j=1$. Each population has a distribution $\Pi_j$ of mortality risk. The distributions $\Pi_j$ are induced by a distribution $M_j$ on covariate $X$ space and transformations $G_j$ from $X$ space to probability space. In practice both $M_j$ and $G_j$ will need to be estimated which we discuss in section \ref{accounting}, but we take them to be known for the current discussion. Let $x_i$ be a random draw from $M_j$; then, for a logistic regression model, functions $G_0$ and $G_1$ would be inverse logit functions $\exp(\cdot) / (1 + \exp(\cdot))$ of $x_i'\beta_0$ or $x_i'\beta_{1}$ where $\beta_0$ and $\beta_1$ are regression coefficients for the reference and comparison populations. Populations $0$ and $1$ may differ in the distributions $M_0$ and $M_1$ of the covariates and in the values $\beta_0$ and $\beta_1$ of the regression covariates that multiply the covariates. 

In comparing $\Pi_0$ to $\Pi_1$, we will make interpretable adjustments to $\Pi_0$ to make it be more similar to $\Pi_1$ so as to understand how and why $\Pi_0$ and $\Pi_1$ differ.

\subsection{Location-Scale Adjusted Distributions}

\label{adjusted}

It is useful to summarize differences between distributions based on a few summary measures. If $\Pi_0$ and $\Pi_1$ were continuous distributions on the real line, we might recenter and rescale $\Pi_0$ so that the resulting distribution $\Pi_A$ has the same mean and variance as $\Pi_1$. We create a new distribution, $\Pi_{A}$ that represents a counterfactual or synthetic population. Then we could summarize the remaining differences between $\Pi_A$ and $\Pi_1$ by a one number summary such as a Kullback-Leibler divergence or other divergence between distributions, or we could plot $\Pi_A$ and $\Pi_1$ and inspect the differences graphically. General location-scale shifts do not work particularly well for distributions with bounded support, though with restrictions they can be useful. Let us assume that $\Pi_0$ and $\Pi_1$ enjoy full support on $(0,1)$. Let $\Pi_A = a + b * \Pi_0$ be a location scale adjusted version of $\Pi_0$ with $a, b \ge  0$. While $b$ could be negative, that flips the distribution around so that large probabilities become small and small probabilities become large, which violates the sense of manipulating one mortality distribution to be more like a second mortality distribution. For $\Pi_A$ to be a distribution contained in $(0,1)$, we need $0 \le a < 1$, $0 < a + b \le 1$, $0 < b \le 1$.

As mortality distributions are generally skewed with a mode close to zero and a long right tail, we prefer to restrict $a = 0$ and $0 < b < 1$. Assuming the mean $\mu_0$ of $\Pi_0$ is greater than the mean $\mu_1$ of $\Pi_1$, we can take $b = \mu_1/\mu_0$. This leaves $\Pi_A$ with support in the range $(0,b) \subseteq (0,1)$. If we think of $\Pi_0$ as the mortality risk in a country at an earlier time point and $\Pi_1$ as mortality risk at a later time, then choosing $b < 1$ is not a problem, as, in the current era, mean infant risk is generally decreasing over time. If mortality risk increased, then we would manipulate $\Pi_1$ instead of $\Pi_0$. Medians could certainly be used in place of means as well and we have used both means and medians in our work.

Thus $b = \mu_1/\mu_0$ summarizes the differences between the distributions $\Pi_0$ and $\Pi_A$. To fully understand the differences between $\Pi_0$ and $\Pi_1$, we would still need to summarize the differences between $\Pi_A$ and $\Pi_1$.

\subsection{Decomposition Methods}

When comparing two distributions, decomposition methods are used to disentangle the effects of differences in coefficients $\beta_0$ versus $\beta_1$ from effects of differences in the distribution of the covariates \citep{Kitagawa:1955, Oxaaca:1973}. For example, mother's age at the birth of an infant is an important covariate for estimating mortality risk as births from younger and older mothers have higher mortality risk. For two populations the relative importance of maternal age may differ and the distribution of maternal age can differ for different populations.

A useful adjustment is to take the distribution of covariates for either population $0$ or $1$ and the regression coefficient $\beta_j$ for the other population. Thus the distribution $\Pi_A$ could be constructed as drawing $x_i$ at random from covariate population $M_1$, then multiplying by coefficient vector $\beta_0$ from population $0$ and taking the inverse logit to construct adjusted distribution $\Pi_A$. A conditional version for sampling $x_i$ is also possible. Suppose that the density of $x_i = (x_{i1},x_{i2})'$ in population $j$ is $f_j(x_i) = f_j(x_{i1}) f_j(x_{i2}|x_{i1})$. We can construct adjusted probability distribution $\Pi_A$ by drawing from a combination covariate density $f_0(x_{i1})f_1(x_{i2} | x_{i1})$, then multiply by $\beta_0$ and taking the inverse logit. This constructs a population of probabilities based on population $\Pi_0$ except that the $x_{i2}$ covariate distribution has been adjusted to look conditionally like covariates from population $1$.

The distributions $\Pi_A$ that we have described here and in the previous subsection are called adjusted or \textit{counterfactual} distributions.

\subsection{Comparing Distributions}

\label{comparing}

We now consider methods to quantify the differences between $\Pi_0$ and $\Pi_1$ by comparing $\Pi_0$ to $\Pi_A$ and also $\Pi_A$ to $\Pi_1$. Let $L$ be a comparison (divergence, distance) operator between distributions that satisfies the triangle inequality
\begin{align}
L(\Pi_0,\Pi_1) \le L(\Pi_0,\Pi_A) + L(\Pi_A,\Pi_1)
\label{triangle}
\end{align}
The Oaxaca decomposition \citep{Oxaaca:1973}, where $L(\Pi_0,\Pi_1) = |\mu_0 - \mu_1|/\mu_1$ is the absolute value of the relative difference in means, follows the triangle inequality when $\mu_0 \le \mu_A \le \mu_1$ where $\mu_A$ is the mean of the adjusted distribution. Similarly, $L(\Pi_0,\Pi_1) = |\mu_0 - \mu_1|$, the absolute difference in means also follows the triangle inequality.

\subsection{Numerical Summaries for Comparing Distributions}

It is often useful to quantify how much one distribution differs from another by a one number summary. Traditional approaches have summarized the distributions first with a one-number summary of the distribution and then compared the summaries. Measures such as the mean, variance, other moments, or quantiles can be used to compare and summarize distributions and we are not opposed to these methods, but we wish to consider other methods as well. We can measure how similar two distributions are by using divergence measures, provided the distributions share the same support. A commonly used measure is the Kullback-Leibler divergence,
\begin{align}
K(\Pi_0,\Pi_{1}) = \int\log\frac{f_0(X)}{f_{1}(X)}f_0(X)\, \mbox{d}X,
\end{align}
where the densities $f_0(\cdot)$ and $f_1(\cdot)$ correspond to the distributions $\Pi_0$ and $\Pi_1$. The divergence $K(\Pi_0,\Pi_{1})$
can be interpreted as the expected information for discriminating $f_{0}(X)$ from $f_1(X)$ based on a single observation from $f_{0}(X)$. Another useful measure is the $L_{1}$ norm
\begin{align}
L_{1}(\Pi_0;\Pi_{1}) = \frac{1}{2}\int\big|f_0(X)-f_1(X))\big| \, \mbox{d}X.
\label{lonenorm}
\end{align}
$L_{1}$ which quantifies how much probability mass needs to be moved so that one distribution becomes identical to the other one \citep{Weiss:1996}. The $L_1$ norm \eqref{lonenorm} is between 0 and 1, with 0 meaning the distributions are the same and 1 meaning the two distributions $\Pi_0$ and $\Pi_{1}$ do not share support.

\section{Using Covariates to Explain Mortality Risk}

\label{covariates}

Variation in mortality risk can often be explained by covariates. By analyzing subpopulations identified by levels of a categorical covariate, we split a large population into several subpopulations. Rather than being concerned with how one subpopulation compares to another subpopulation when we have a set of $K$ subpopulations, we are more concerned with how distinct the subpopulations are and whether targeting one or anther subpopulation for intervention might be useful for reducing overall infant and child mortality  risk. However, we have found that even covariates that explain significant amounts of risk are not necessarily very useful in identifying subgroups to target so as to alleviate high risk infant and child mortality , a situation we illustrate in section \ref{modeling}.

\subsection{ANOVA Methods}

Consider a categorical covariate $X$ that identifies groups, for example wealth quintiles or maternal age groups. If most of the variability in risk occurs between groups, then knowing the mean risk for each group identified by $X$ is highly informative about inequality in the population. In contrast, if most of the variability in risk occurs within groups, then comparing mean risk between groups does not provide much information on risk variation.

Variation in risk can be expressed as the between group variance plus the weighted sum of the within-group risk variances. The law of total variance produces the decomposition
\begin{align}
\Var(\Pi_0)&= \EEE_{X}\left( \Var[\Pi_0|X]\right) + \Var_{X}\left[ \EEE(\Pi_0|X) \right]
\end{align}
where $\EEE_{X}(\Var[\Pi|X])$ is the average within group variance and $\Var_{X}[\EEE(\Pi_0|X)]$ is the between-group variance of the group means. We fit Analysis Of Variance (ANOVA) models using mortality risk as outcome and group membership as predictor and use $R^{2}$ as a measure of how much of the total variance in infant and child mortality  risk can be explained by membership in a particular group.

As an example, in section \ref{modeling} we run separate ANOVAs for covariates and for births from a given year in India. We consider whether the $R^2$ might be increasing or decreasing over the years, to identify whether inequality across covariate categories might be increasing or decreasing with time. 

It is not necessary that the ANOVA model be as complex as the Bayesian hierarchical model used to model the data. We consider the ANOVA as a summary measure of the posterior, not a model on its own right. Summarizing the results of the hierarchical model with a single covariate is useful, even if the hierarchical model has several covariates and interactions, because public health interventions often use a single covariate to determine eligibility  to policy interventions. 

\subsection{Time Trends}

We are particularly interested in time trends in infant and child mortality  and how group comparisons evolve over time in India. We use year to define subpopulations, and then use additional covariates such as income quintiles to further refine the population of births into sub-sub-populations. We use ANOVA to quantify the within-year inequality within and across income groups and then assess the trend in this summary over time. Among other additional comparisons, we use divergence measures to compare the distribution of mortality in the baseline year against the distribution of mortality risk in each of the following years. We can construct synthetic populations holding the distribution of some key covariate constant to evaluate its effect on mortality risk time trend. For example, we can estimate the distribution of mortality risk over time if the distribution of maternal age was fixed over time.

\section{Accounting for Uncertainty When Evaluating Inequality}

\label{accounting}

We estimate risk for subjects with a Bayesian hierarchical logistic random effects regression models fit using Markov chain Monte Carlo (MCMC). For each iteration $\ell = 1, \ldots, L$ of the MCMC  , we have a point estimate $\pi_i^{(\ell)}$ of $\pi_i$ for all births and we can now proceed with our program to evaluate infant and child mortality  risk inequality as sketched in sections \ref{analyzing} and \ref{covariates}. For example, we can use the $\pi_i^{(\ell)}$s and each birth's wealth quintile $x_i$ to calculate an $R^{2\! (\ell)}$. However, this calculation is for only a single iteration of the MCMC. Thus for all iterations $\ell = 1, \ldots, L$ we calculate an $R^{2\! (\ell)}$ producing a posterior distribution for $R^2$. This uncertainty can be used in comparing various $R^2$s for different covariates and thus determine which covariates do the best at distinguishing infant and child mortality  risk.

As our sample $\pi_i^{\ell}$ is discrete, when we wish to calculate the divergence between two subpopulations of $\pi_i^{\ell}$s, we use a kernel density smoother to first smooth out the distributions before calculating a divergence. Then, as with ANOVA $R^2$s, we can calculate distributions for divergences.

In our graphs and numerical summaries we present point estimates and 95\% pointwise credible intervals calculated from the multiple MCMC samples.

\section{Modeling of Early Life Mortality Risk in India}

\label{modeling}

Our data on infant mortality in India comes from the Demographic and Health Surveys, (DHS) \url{http://www.measuredhs.com/}. We use data from two DHS surveys to construct a \emph{restrospective panel} from 1975 to 1997. A third wave covering more recent years is available but does not include district level information which we use in our models, so we were unable to make use of the third wave data. We analyze births to mothers aged 15--35 from 1975 through 1998 to reduce truncation and censoring. We analyze a total of 408,706 births from 141,999 mothers in 3,806 sampling clusters taken from 443 districts and 26 states.


\subsection{Model Specification and Estimation of Mortality Risk}

We fit a hierarchical Bayesian logistic regression model to estimate each infant $i$'s mortality risk $\pi_{i}$ using covariates, time, and time varying covariate effects.

Let $i = 1, \ldots, N$ index the $N$ births, nested in $m = 1, \ldots, M$ mothers, nested in sampling clusters $c = 1, \ldots, C$, nested in districts $d = 1, \ldots, D$, nested in states $s = 1, \ldots, S$. Year of birth is indexed by $t = 1, \ldots, T$ with $t=1$ for births from 1975 up to $t=23 \equiv T$ for births from 1997. Let $y_{i}$ be the binary response variable, whether birtht $i$ with covariate vector $x_i$ born from mother $m(i)$, in sampling cluster $c(i)$, in district $d(i)$, in state $s(i)$, and in year $t(i)$ was alive $y_{i} = 0$ at the age of one or not $y_{i} = 1$. We specify random intercepts for mother and cluster and bivariate random intercepts and time slopes at the district and state level. Let $\logit(\pi) = \log(\pi/(1-\pi))$. Our Bayesian hierarchical logistic regression is
\begin{align}
y_{i}| \pi_{i} & \sim \Bern(\pi_{i}) \\
\logit(\pi_{i}) &= x'_{i}\alpha + \delta_{m(i)} + \gamma_{c(i)} +\xi_{d(i),1} + \xi_{d(i),2}*t(i) + \tau_{s(i)1} +
\tau_{s(i)2}*t(i),
\end{align}
where $\pi_{i}$ is the unobserved probability of mortality for infant $i$, $\alpha$ is a vector of unknown coefficients corresponding to the covariates in $x_{i}$, $\delta_{m}$ is the mother random effect with variance $\sigma^{2}_{1}$, $\gamma_{c}$ is the sampling cluster random effect with variance $\sigma_{2}^{2}$, $\xi_{d,1}$ and $\xi_{d,2}$ are district random intercepts and slopes with prior covariance matrix $\bfSigma$, and $\tau_{s1}$ and $\tau_{s2}$ are state random intercepts and slopes with prior covariance matrix $\bfPsi$. Define $\xi_d = (\xi_{d1}, \xi_{d2})'$ and $\tau_s = (\tau_{s1},\tau_{s2})'$. The random effects priors are $\delta_{m} | \sigma^{2}_{1} \sim \NNN(0,\sigma^{2}_{1})$, $\gamma_{c} | \sigma^{2}_{2} \sim \NNN(0,\sigma^{2}_{2})$, $\xi_{d} | \bfSigma \sim \NNN_2(\bfzero, \bfSigma)$, and $\tau_{s} |\bfPsi \sim  \NNN_{2}(\bfzero,\bfPsi)$.

Covariates at the child level are birth order, birth year, gender, maternal age, with household level covariates of religion, caste, wealth quintile, residence (rural or urban), and maternal education. We use splines to capture non-linearities in the time trend and investigated whether covariates have time-varying effects, though these analyses are not shown in this paper. We include main effects and all $_9C_2 = 36$ two-way interactions between covariates. The effects for maternal age, birth order, and birth year are modeled using b-splines. 

The intercept is given a $\text{N}(0,9)$, prior, all main effects are given standard normal priors, and all two-way interactions are given $\text{N}(0,0.5)$ priors. The web appendix gives the complete model specification.

While interest in individual probabilities $\pi_{i}$ is standard in hierarchical logistic regressions, we are not really interested in individual probabilities. Rather, we are interested in the entire collection of probabilities $\pi_{i}$, $i = 1, \ldots, N$ simultaneously as the key quantity of interest from the model. This set of probabilities are used as inputs to our inequality calculations. We generate samples from the posterior distribution of the infant mortality risks, $\pi_{i}^{(\ell)}, i = 1, \ldots, N$, and where $\ell = (1, \ldots, L)$ indexes the $L$ MCMC samples.

\subsection{Analysis of Inequality}

\begin{figure}[H]
	\centering
	\includegraphics[height=8cm,width=14cm]{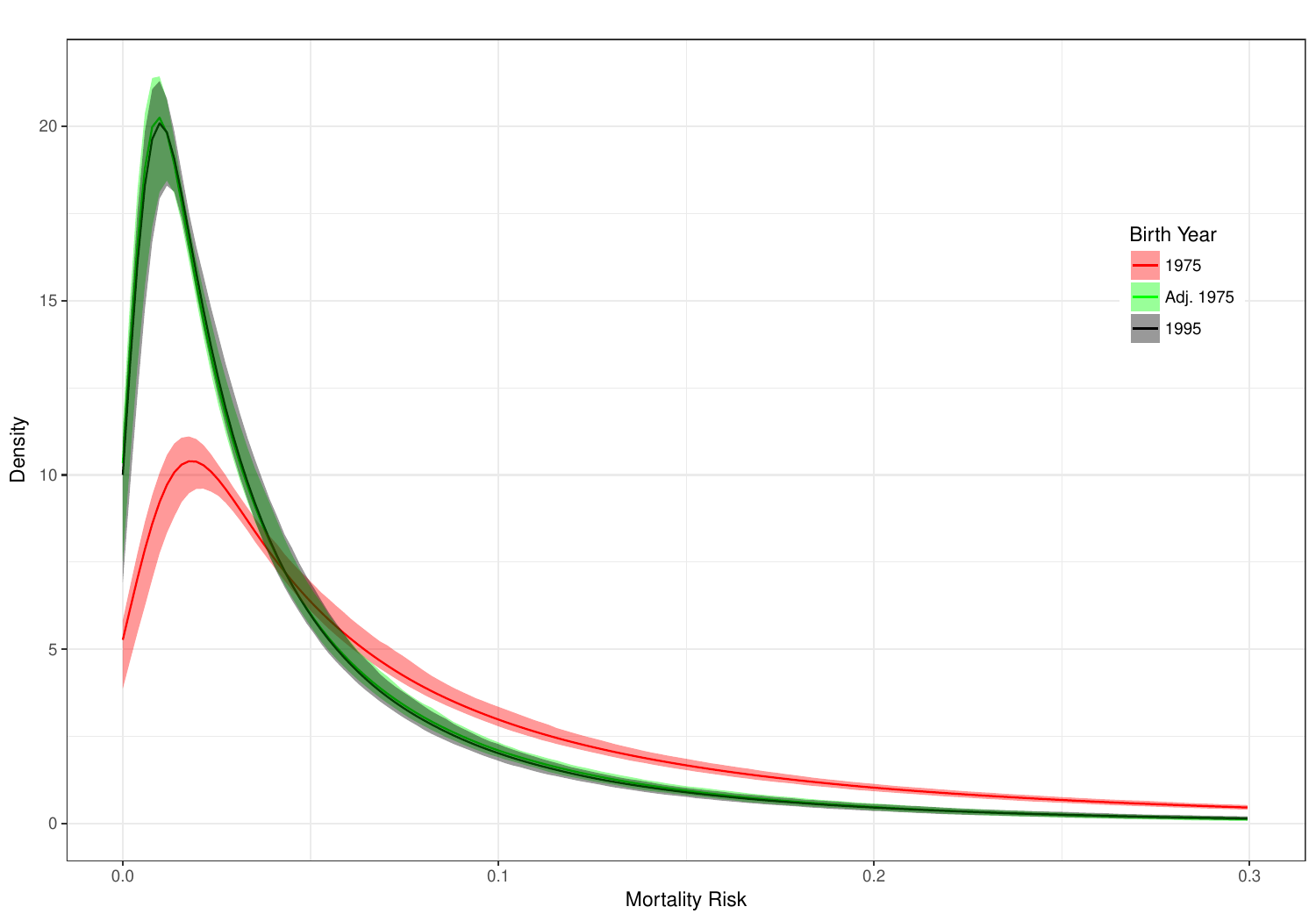}
	\caption{Kernel densities of mortality risk for 1975 and 1995 and a counterfactual adjusted 1975 population. Intervals around the main curve are 95\% pointwise posterior bounds for the density height. The counterfactual $1975_A$ population takes the values from 1975 and multiplies by the ratio of the median risks from 1995 to 1975. The differences between 1975 and 1995 appear to be fully summarized by the median adjustment because it is almost impossible to distingush the gray live from the green line.}
	\label{adjusteddistri}
\end{figure}

Disparities in mortality risk are not well-captured by looking at national averages of mortality rates. For example, the infant mortality rate (IMR) in India was 12\% in 1975 and 6\% in 1995, both calculated as unadjusted means from our data. While this is a remarkable decline, these numbers do not quantify important aspects of the distribution of mortality risk as estimated by our model. To illustrate, figure \ref{adjusteddistri} displays three kernel density estimates: one representing the distribution of mortality risk in India for 1975, which is the less peaked, longer right tailed density, and one for 1995, which has the higher mode and smaller right tail and a third density mostly hidden next to the 1995 density which we discuss in a moment. The shaded areas display 95\% pointwise credible regions; we generated $L$ kernel density estimates from $L$ MCMC samples $\pi_i^{(\ell)}$ and calculate the .025 and .975 percentiles at each point along the $x$ axis. The modal mortality risk was 2\% in 1975 and 1\% in 1997, which means that the actual mortality risk most infants experienced in both years were quite different from the IMR. In 1975 26\% of infants have mortality risk higher than the 12\% 1975 IMR while in 1995 only 11\% of infants have mortality risk higher than the 1975 IMR of 12\%. The uncertainty in the kernel density estimates shows that the distribution of infant mortality risk from our model is well determined in both years, and thus we are comfortable in asserting that the mortality risk distribution is stochastically much smaller in 1995 than in 1975. The third density plotted in Figure \ref{adjusteddistri} is almost indistinguishable from that of 1997 and was constructed by taking the probabilities in 1975 and multiplying by the ratio of the 1997 median to the median for 1975. The resulting adjusted 1975 density is virtually indistinguishable from the 1997 distribution. Summary statistics of the densities in figure \ref{adjusteddistri} are given in table 2.

\begin{figure}[H]
	\centering
	\includegraphics[height=8cm,width=14cm]{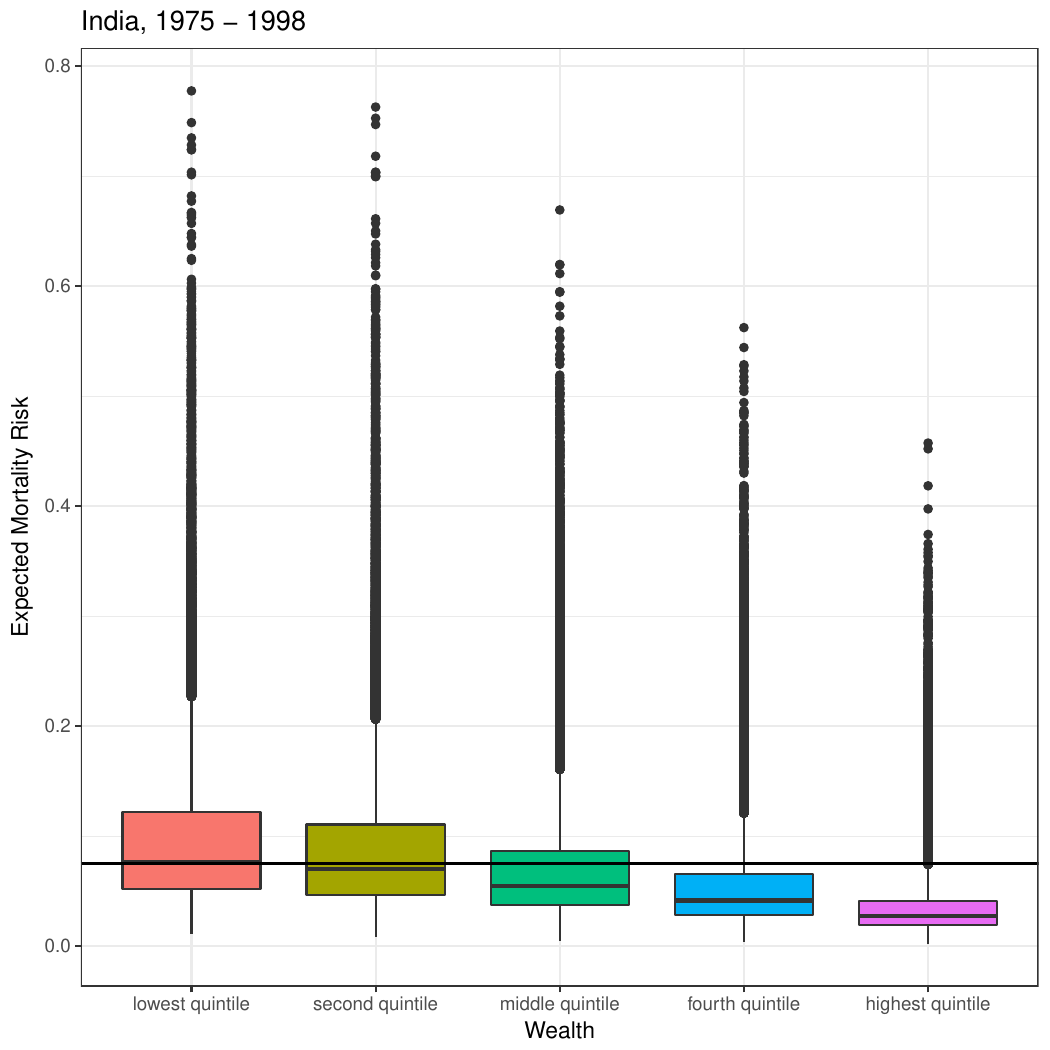}
	\caption{Distribution of mortality risk by wealth quintile in India. Each box plot summarizes the distribution of individual infant average mortality risk $\EEE[\pi_i|Y]$ in each wealth quintile.}
	\label{wealthquintile}
\end{figure}

Looking at the median of $\pi_i$ by wealth quintile we find that the  mortality rate is: lowest quintile .08; second .07; middle .05; fourth .04; highest .03, which shows that the mortality rate is almost 3 times higher than the poorest group compared to the richest. However, these calculations ignore within-group variability. Figure \ref{wealthquintile} shows box plots of posterior means $\EEE[\pi_{i}|Y]$ of mortality risk for every infant in our data, by wealth quintile, where $Y$ denotes the entire data set. By contrast with the means by quintile, figure \ref{wealthquintile} shows that the distribution of mortality risk is highly variable within each wealth quintile. For example, 10\% of births from the two lowest quintiles have mortality risk higher than 12\% and 30\% of births from other wealth quintiles have higher mortality risk that the $.08$ median mortality risk among the poor. This suggests that summaries such as means hide useful information and that quantifying individual mortality risk is necessary.

\begin{figure}[H]
	\centering
	\includegraphics[height=8cm,width=15cm]{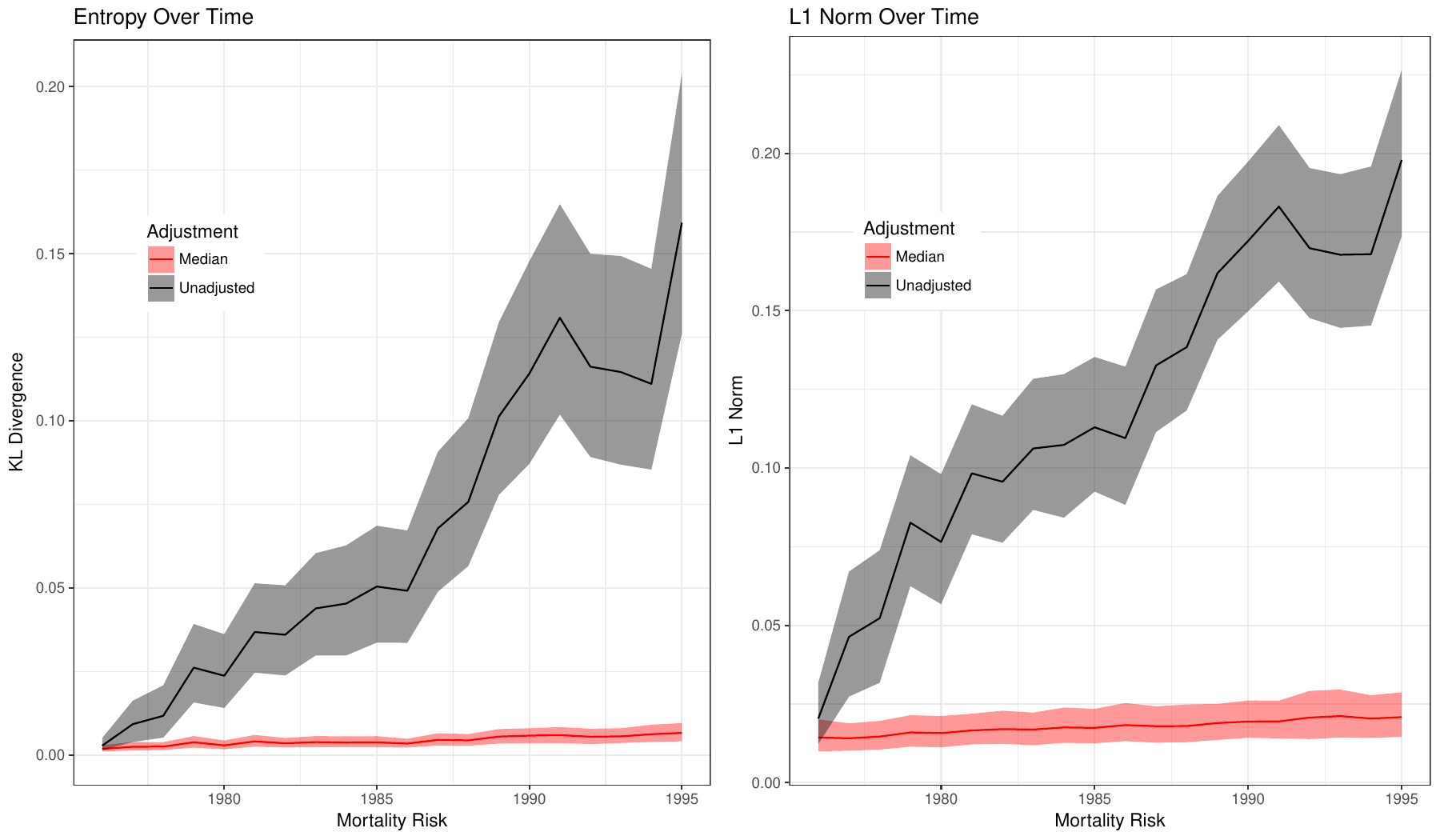}
	\caption{Inequality trends in infant mortality in India, 1975--1995, with a multiplicative median ratio adjustment in the relative distribution. The baseline year is 1975 and all subsequent years are compared against the baseline year.}
	\label{multiplicative}
\end{figure}

We quantify how much of the changes in infant mortality over time can be explained by the multiplicative adjustment. We use 1975 as the comparison year to compare against each subsequent year $1976, \ldots, 1995$. In figure \ref{multiplicative} we use KL (left) and $L_1$ (right) divergence measures to compare differences in the distributions year $t$ versus 1975, with (lower dotted curve) and without (upper solid curve) the multiplicative median adjustment. Bands are pointwise 95\% posterior intervals. The differences between the baseline year and the other years are close to zero after the adjustment and do not grow larger with time. The overall distribution of mortality risk in India grows increasingly different from the baseline distribution in 1975, yet these changes can mostly be explained by a parsimonious multiplicative shift. This suggests that the primary change in risk is due to improvements in the intercept and any changes in covariates had countervailing changes in other covariates.

\begin{figure}[H]
	\centering
	\includegraphics[height=8cm,width=14cm]{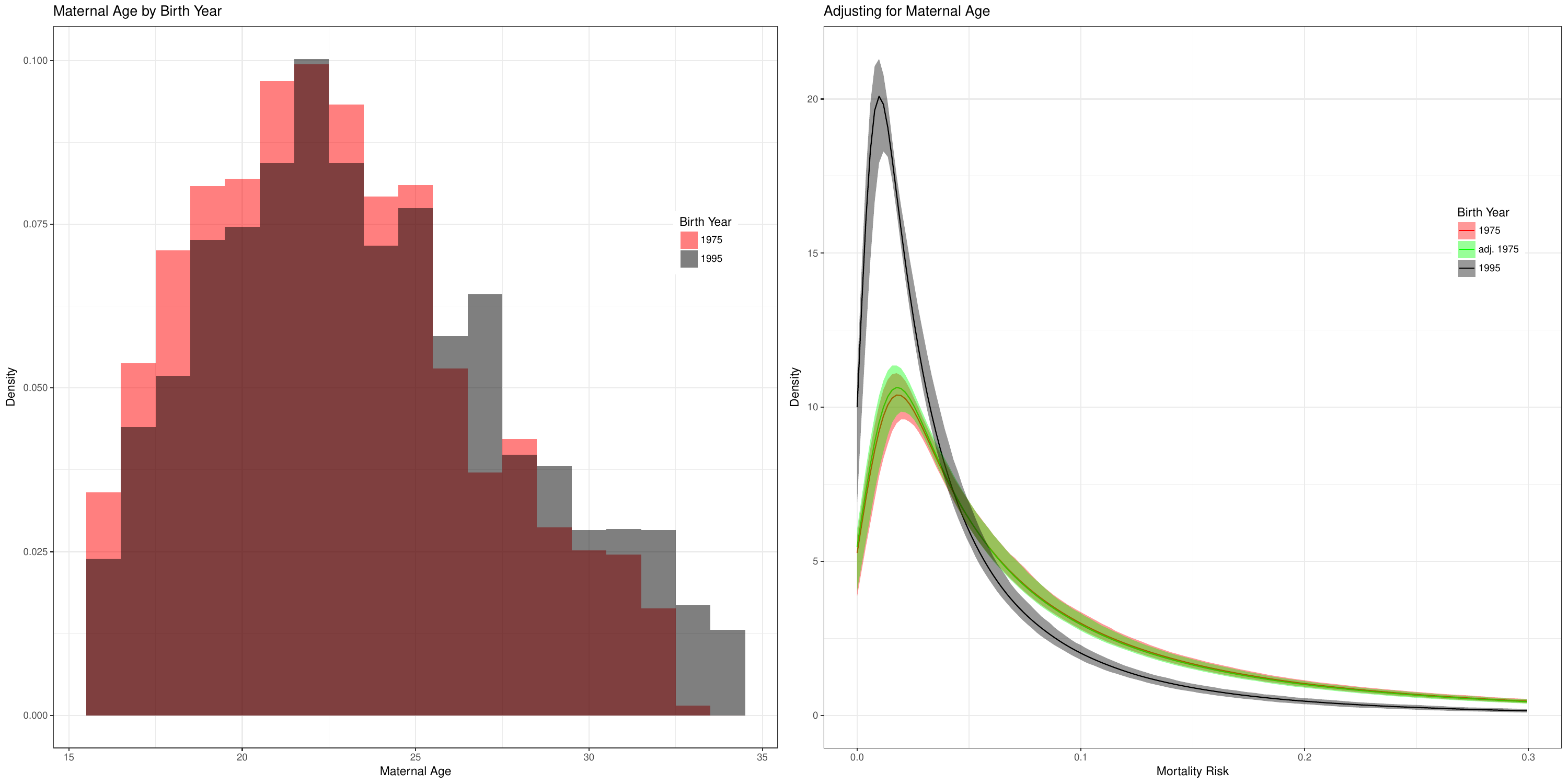}
	\caption{Distribution of maternal age and its effects on the distribution of mortality risk for 1975 and 1995. Graph on the left is a histogram with the distributions of maternal age for 1975 (pink) and 1995 (grey). The graph on the right represent three densities of mortality risk, for 1975 (green), for 1995 (pink), and 1975 adjusted to have the same distribution of maternal age that 1995 (green). Most of the differences in the distribution of mortality risk between these two years cannot be explained by the change in the distribution of maternal age.}
	\label{maternalage}
\end{figure}

Figure \ref{maternalage} analyzes the effect of the change in maternal age distribution over time on the distribution of mortality risk. The left panel plots histograms of maternal age in 1975 (pink) and 1995 (grey). While there is a great deal of overlap in the age densities, maternal age in 1995 has fewer younger mothers as compared to 1975. The right-hand panel displays the distribution of infant mortality risk for 1975 and 1995 and an adjusted version of 1975 where we have adjusted the maternal age distribution to look like that in 1995. Adjusting for maternal age in 1975 changes the adjusted 1975 distribution slightly towards smaller values, but not far enough. Table 3 presents Kullback-Leibler and $\text{L}_{1}$ divergences quantifying the differences between the adjusted 1975 distributions of infant mortality and the 1995 (unadjusted) distribution after adjustments for the various covariates in our sample with 95\% posterior intervals in parentheses. Single covariate adjustments alone do not explain the changes in distribution of mortality risk from 1975 to 1995.

\begin{figure}[H]
	\centering
	\includegraphics[height=10cm,width=16cm]{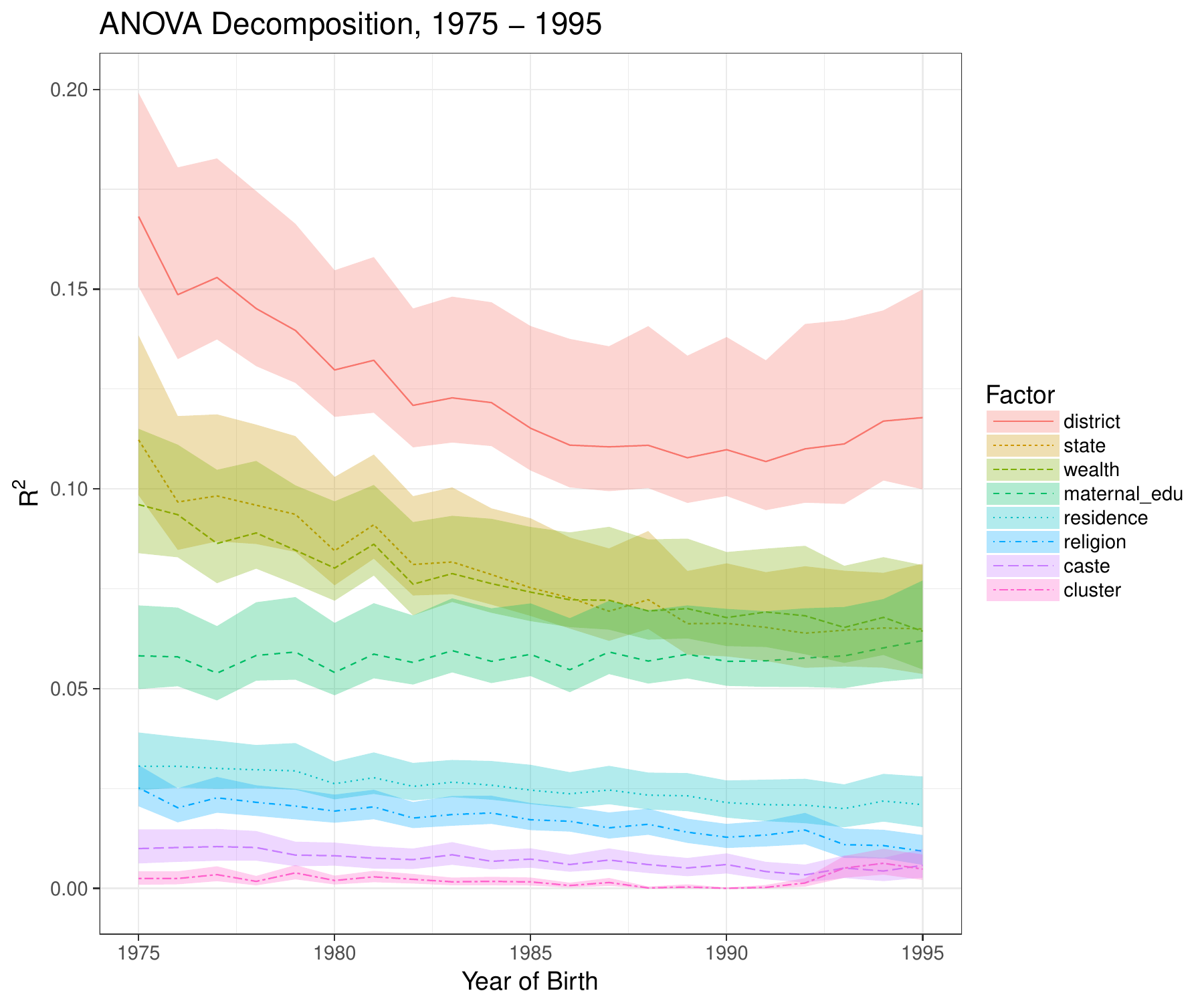}
	\caption{Fraction of mortality risk explained by individual covariates for India over time, 1975--1997. Each line represents the trend for a particular covariate, showing how much of the variability in mortality risk is due to between-group sources. The shaded areas represent 95\% pointwise credible intervals.}
	\label{trends}
\end{figure}

Figure \ref{trends} shows the extent to which variability in mortality occurs within covariate categories and doesn't change over time. We look at the $R^{2}$ for caste, religion, states, maternal education, wealth, residence, and districts over time. For all covariates and all times, the vast majority of variation in inequality in infant mortality is within categories and not between categories. For religious group and for caste, the covariate contains almost no information on the variability of inequality: for these groups, almost 100\% of the variation is within category. These patterns are consistent over time. This suggests that socioeconomic groups described by a single covariate are much more heterogeneous than previously thought and that infants within a single sub-group defined by levels of wealth or other covariates have very different mortality risks. This also confirms the notion that comparisons between the means of groups defined by levels of a categorical demographic variables ignores substantial within-group variability in mortality risk due to the high levels of heterogeneity. This finding highlights the importance of measuring inequality within as well as between groups of births. This is also an important contribution to the literature on India and it has been overlooked by previous research.

\section{Discussion and Conclusions}

\label{discussion}

Our findings have important implications for health policy and inequality measurement. Currently inequality in early-life mortality often uses either measures from the income inequality literature or calculates average differences (or ratios) between large groups of births, such as between countries or between income groups within-countries. Our results show that these analyses are incomplete, and possibly misleading. Because national averages of child mortality are declining over time, using measures from the income inequality literature to track changes over time in inequality in mortality can cause the false impression that inequality is increasing around the globe. As we have discussed, because all these measures have population average in the denominator, as in equation \ref{inequality}, and thus we are dividing the formula by an increasingly smaller number, the overall inequality measures may be increasing over time. Similarly, income inequality measures will provide inconsistent answers whether we are quantifying mortality or survival. That might the reasons a recent report on India shows that Gini and similar indices applied to infant and child mortality are increasing over time \citep{India:2012}. 

Alternatively, quantifying mortality by calculating average differences in mortality rates between large groups of births will miss the within group component of variability in mortality risk. For example, several studies have suggested that some Sustainable Development Goals (SDG) were not achieved due to high levels of inequality within countries \citep{India:2012, Houweling:2009, Gwatkin:2005}. UN General Assembly Resolution 68/261, which highlights Sustainable Development Indicators as key measurements for measuring progress in reducing early-life mortality, recognizes this fact and recommends that health inequality measures should be disaggregated, where relevant, by demographic groups \citep{UN:2016}. Our findings support the UN concerns but go even further. They suggest that even monitoring inequality between-groups of births may not be enough to identify left behind children because there are a considerable number of births that have higher mortality risk than the reported group averages. Thus to design effective policy interventions depends on quantifying mortality risk at a much more fine grained level and evaluating inequality within and between group of births. This study supports the view that monitoring inequality across births is useful for policy purposes \citep{Murray:2001,Gakidou:2002,Braveman:2001,Ramos:2018}.

Finally, our results have important implications for program targeting. Health policies often target births based on a single risk factor, most commonly poverty, for simplicity and because  infants in poor households have higher average mortality rates compared to infants in richer households \citep{Houweling:2009,Gwatkin:2004,Black:2003,Braveman:2001,BRYCE:2003,Jones:2003,Victora:2003}. This approach assumes that most births within the same group have similar mortality risk. The most common interventions that target births based on poverty are perhaps Cash Transfer Programs (CTP), now widely implemented in many low and middle income countries \citep{Glassman:2013,Bassett:2008,Akresh:2015} but there are many other types of child health interventions that target births from poor families \citep{Banerjee:2010,Huicho:2016,Gakidou:2007} While births from  the poorest families have higher mortality rates than other births, targeting births based solely on poverty --- or based on any other single risk factor --- ignores within group heterogeneity, where births from the same group may have very different mortality risks \citep{Gakidou:2002,Gakidou:2003,WHO:2000,India:2012} . As discussed by \cite{Ramos:2018}, targeting groups that are highly heterogeneous in mortality risk is inefficient for program targeting because it allocates resources to lower risk births not in need of program resources and miss many high risk births that need interventions. 

While in this paper we have looked at infant and child mortality , our results extend to mortality data in general, such as adult mortality. Because mortality risk is a probability, it is fundamentally different from income and our results suggest that measures from income inequality literature are inappropriate for mortality. More generally, our methods have broad applicability to other health outcomes, including those that are not defined on the probability scale, such as life expectancy or malnutrition. Our methods are particularly useful when scientists suspect that within-group variability can be substantial or when researchers are interested in aspects of the distributional differences between two populations besides mean differences. For example, our methods can be used to determine if the changes over time in height among children is driven by relatively slow growth of certain high-risk children versus faltering of the entire population. 

Although we use a Bayesian approach in this paper, our methods are potentially compatible with a frequentist approach. If a researcher can fit a frequentist model and simulate predictions for mortality risk, for example using bootstrap methods, our methods can be applied to the bootstrap predictions. One notable advantage of the Bayesian approach is that is makes inference easier. For example, the popular LASSO \cite{Tibshirani:1996} can do variable shrinkage and selection, but does not naturally provide standard errors for coefficient estimates and so also won't provide uncertainty measurements for probabilities which are functions of the coefficient estimates. By contrast, in a Bayesian model, we can easily calculate the probability that one sub-population is more unequal than another sub-population by counting the proportion of MCMC samples in which one sub-population's measure of inequality is higher than another. The frequentist approach can depend on being able to fit a complex model to bootstrap samples, something that can fail, for example in attempting to bootstrap a logistic regression with a small fraction of cases or where a variance component gets set to zero when using maximum likelihood.

\clearpage
\appendix

\section{Priors}\label{app}
We use gamma priors on the precision parameters. For the mother random effects,
\begin{align*}
\text{M}_j &\sim \text{N}(0, \sigma_{\text{M}}^{2})\\
\sigma_{\text{M}}^{2} &\sim \text{Inverse Gamma}\left(3, 2\right)
\end{align*}
Similarly, for clusters,
\begin{align*}
\text{C}_l &\sim \text{N}(0, \sigma_{\text{C}}^{2})\\
\tau_{\text{C}}^{2} &\sim \text{Inverse Gamma}\left(3,2\right),
\end{align*}
For districts,
\begin{align*}
\begin{pmatrix}
\text{D}_{k,1} \\ \text{D}_{k,2}
\end{pmatrix} &\sim \text{N}_2 \left(
\bm{0}_{2 \times 1},
\bm{\Sigma}_{\text{D}}
\right) \\
\bm{\Sigma}_{\text{D}} &\sim \text{Inverse Wishart} \left(\begin{pmatrix}
1 & 0 \\ 0 & 0.1
\end{pmatrix},4\right)
\end{align*}
For states,
\begin{align*}
\begin{pmatrix}
\text{S}_{m,1} \\ \text{S}_{m,2}
\end{pmatrix} &\sim \text{N}_2 \left(
\bm{0}_{2 \times 1},
\bm{\Sigma}_{\text{S}}
\right) \\
\bm{\Sigma}_{\text{S}} &\sim \text{Inverse Wishart} \left(\begin{pmatrix}
1 & 0 \\ 0 & 0.1
\end{pmatrix},4\right)
\end{align*}

All fixed effects are given mean zero normal priors. The intercept term has variance 9, all other main effects have variance 1, and two-way interactions have variance 0.5. In total, there are
\begin{align*}
\sum_{h = 1}^2 \begin{pmatrix}
9 \\ h
\end{pmatrix} + 1 = 46
\end{align*}
covariates. The categorical covariates can have differing numbers of levels, so this gives a total of $6,505$ regression parameters to be estimated.

\clearpage

\bibliography{measurement_inequalities,antonio.books}
\bibliographystyle{Chicago}

\clearpage

\begin{table}[h]
	\centering
	\label{betasimulation}
	\scalebox{0.8}{
		\begin{tabular}{r rr rr rr rr rr rr}
			\hline
			Beta & \multicolumn{2}{c}{Mean}   & \multicolumn{2}{c}{SD}  & \multicolumn{2}{c}{CV}  & \multicolumn{2}{c}{Gini}  & \multicolumn{2}{c}{Theil}  \\
			
			$(\alpha,\beta)$ & Mort & Surv  & Mort & Surv & Mort & Surv & Mort & Surv & Mort & Surv \\ \hline
			(1, 10) & 0.0909 & 0.9091 & 0.0829 & 0.0829 & 0.9126 & 0.0912 & 0.4761 & 0.0476 & 0.3778 & 0.0044 \\ 
			(0.5, 10) & 0.0476 & 0.9524 & 0.0628 & 0.0628 & 1.3193 & 0.0660 & 0.6210 & 0.0311 & 0.6831 & 0.0023 \\ 
			(0.3, 10) & 0.0291 & 0.9709 & 0.0500 & 0.0500 & 1.7160 & 0.0515 & 0.7207 & 0.0216 & 0.9864 & 0.0014 \\ 
			(0.1, 10) & 0.0099 & 0.9901 & 0.0297 & 0.0297 & 3.0022 & 0.0300 & 0.8787 & 0.0088 & 1.8303 & 0.0005 \\ 
			\hline
		\end{tabular}
	}	
\caption{Income inequality measures applied to simulated $\betadist(\alpha,\beta)$ infant and child mortality distributions. The first column $(\alpha,\beta)$ gives the parameters of the beta distribution.  After the first column, each pair of columns gives a particular measure applied to the probability of Mortality (Mort) distribution and the probability of Survival (Surv) distribution. Pairs of columns are the mean, the standard deviation (SD), coefficient of variation (SD/mean), Gini and Theil inequality measures. }
\end{table}

\clearpage
\begin{table}[H]
	\centering
	\label{sometable}
	\begin{tabular}{rll}
		\hline
		& 1975 & 1995 \\
		\hline
		IMR &  $12.0\%$& \hspace{0.6em}$6.0\%$ \\
		Mode & \hspace{0.5em}1.8\% (1.6\%, 2.3\%) & \hspace{0.5em}1.0\% (0.9\%, 1.3\%) \\ 
		Mean & \hspace{0.5em}9.6\% (9.2\%, 10.1\%) & \hspace{0.5em}5.2\% (5.0\%, 5.4\%) \\ 
		Median & \hspace{0.5em}5.4\% (5.1\%, 6.0\%) & \hspace{0.5em}2.8\% (2.6\%, 3.1\%) \\ 
		Percent infants with \\ $\pi_i>$  1975 IMR & 25.7\% (24.2\%, 27.3\%) & 10.8\% (9.6\%, 11.7\%) \\     \hline
	\end{tabular}
	\caption{Summary statistics of the densities in Figure \ref{adjusteddistri}. The infant mortality rate (IMR) is the empirical proportion of births that experienced mortality and is calculated from the raw data.}
	
\end{table}

\clearpage

\begin{table}[H]
	\centering
	\begin{tabular}{rcc}
		\hline
		Covariate & Entropy & L1 Norm \\ 
		\hline
		Overall & 0.182 (0.144, 0.230) & 0.278 (0.225, 0.359) \\ 
		Maternal Education & 0.130 (0.100, 0.168) & 0.229 (0.181, 0.294) \\ 
		Maternal Age & 0.169 (0.133, 0.214) & 0.267 (0.215, 0.342) \\ 
		Birth Interval & 0.177 (0.141, 0.225) & 0.275 (0.222, 0.353) \\ 
		State & 0.177 (0.141, 0.223) & 0.275 (0.222, 0.355) \\ 
		Religion & 0.178 (0.142, 0.226) & 0.276 (0.223, 0.355) \\ 
		Gender & 0.182 (0.144, 0.230) & 0.278 (0.225, 0.359) \\ 
		Caste & 0.185 (0.148, 0.233) & 0.280 (0.226, 0.362) \\ 
		Birth Order & 0.185 (0.148, 0.234) & 0.280 (0.227, 0.362) \\ 
		Residence & 0.196 (0.157, 0.245) & 0.289 (0.234, 0.370) \\ 
		Wealth & 0.234 (0.191, 0.289) & 0.318 (0.259, 0.407) \\ 
		\hline
	\end{tabular}
		\caption{The effect of various single covariate adjustments to 1975 to see how close the distribution of infant and child mortality then gets to the distribution in 1995, as measured by the $L_1$ norm and entropy. Numbers in parenthesis are 95\% posterior intervals.}
	\label{multipleadjustments}

\end{table}

\clearpage

\clearpage

\clearpage

\clearpage

\clearpage

\end{document}